# Process-Technology Co-optimization for 2D-FETs

S-H. Yang[1,2,†], J. Shah[1,2,†], H. Jawa[1,2], M. Das[3], Y. Tan[1,2], H-Y. Lan[1,2], H-C. Chien[1,2], S. Tripathi[3], M. Villena[5], X. Wu[4], D. Cott[4], K. Banerjee[4], P. Morin[4], C.J.L. de la Rosa[4], G. Thareja[6], D. Lin[4], J. Appenzeller[1,2], Z. Chen[1,2,*]

[1]Birck Nanotechnology Center, Elmore Family School of Electrical and Computer Engineering, [2]Purdue University, West Lafayette, USA, [3]Imec USA, West Lafayette, Indiana 47901, USA, [4]Imec, 3001 Leuven, Belgium, [5]Department of Electronics and Computer Technology, University of Granada, Spain, [6]Applied Materials, CA, USA. [†]Equal contribution. [*]Email: zhchen@purdue.edu

*Abstract*— We present the first experimental machine learning (ML)-enabled Process-Technology Co-Optimization (PTCO) framework for optimizing 2D transition metal dichalcogenide (TMD) FET fabrication directly from statistically meaningful experimental data rather than pure simulation data. We first introduce a transition voltage metric, $V_{\text{Trans}}$, to quantify the gate voltage required for off-to-on switching and reveal its direct correlation with subthreshold swing (SS), highlighting an overlooked switching characteristic that governs both off-state and on-state performance. By integrating automated metric extraction, multi-objective recipe ranking, and predictive modeling, our framework uncovers hidden process-performance correlations and predicts the performance of unexplored fabrication recipes from limited experimental data. Experimental validation shows close agreement with ML predictions, thus demonstrating the framework's ability to efficiently guide gatestack optimization through iterative experimental feedback.

## I. Introduction

2D TMDs are promising candidates for future logic technologies because of their high performance, atomically thin bodies and back-end-of-line compatibility. However, their strong sensitivity to gate-stack formation, interfacial quality and process-induced variability makes experimental optimization intensive while the underlying process–performance relationships remain poorly understood. Existing ML-based DTCO/STCO approaches primarily rely on simulation-generated datasets (TCAD, BSIM etc.), which are effective for mature technologies but do not capture experimentally induced interface effects and variability in emerging material systems. To address this gap, we introduce PTCO, an experimental data-driven framework that learns process–performance relationships from statistically meaningful device populations. Automated analysis extracts SS, $I_{\text{ON}}$ and $V_{\text{Trans}}$ from thousands of measured transfer characteristics. These metrics are integrated into a multi-objective ranking framework and coupled with Gaussian Process Regression (GPR) to identify process–performance correlations and predict unexplored fabrication recipes within the experimental process space. PTCO thereby prioritizes promising recipes, reduces the search space and improves iteratively through experimental feedback.

## II. Transition Voltage and Clustering

### *A. Motivation for Defining Transition Voltage*

Transistor performance is conventionally evaluated using $I_{\text{ON}}$, ON/OFF ratio and SS, which adequately describe mature Si MOSFETs but do not fully capture the switching behavior of emerging 2D FETs. As shown in Fig. 1(a), many high performance 2D devices with competitive SS and $I_{\text{ON}}$ still require a much larger gate voltage window to transition from the off-state to the deep on-state than state-of-the-art Si MOSFETs. This broadened transition increases the required operating gate voltage and represents an overlooked switching characteristic not captured by conventional metrics. Its physical origin in 2D TMD FETs remains unclear but possible mechanisms include process-induced trap charges, distributed interface and border traps, gate-dependent Schottky barrier modulation with near-band-edge trap states at the source contact and other process-dependent interface effects that weaken gate control in the transition region [2]. To illustrate these effects, Fig. 1(c) presents Ginestra™ simulations [3] showing that defects introduced in the 2D TMD channel or at different interfaces degrade SS and broaden the off-to-on transition, consistent with the experimentally observed behavior.

### *B. Transition Voltage Extraction*

For each measured transfer characteristics (example shown in Fig. 2), we identify the onset of subthreshold regime ($V_{\text{G,ssstart}}$), the point of minimum SS ($V_{\text{G,ssmin}}$) and the onset of the near on-state curvature ($V_{\text{G,knee}}$) where $\frac{d\log_{10}(I_D)}{dV_G}$ remains below a half-Gaussian threshold of 35% of that at $V_{\text{G,ssmin}}$ for consecutive points. Then, $V_{\text{G,mid}}$ was taken as the midpoint in sampled voltage index space between $V_{\text{G,ssmin}}$ and $V_{\text{G,knee}}$, and finally the transition voltage was calculated as $V_{\text{Trans}} = V_{\text{G,knee}} - V_{\text{G,mid}}$ to quantify the subthreshold-to-on-state transition.

### *C. Similarity Cluster Analysis*

Fig. 3 shows the extracted $V_{\text{Trans.}}$ versus SS for devices with the same test vehicle design but fabricated using different gate stack process recipes. Fig. 4 shows the corresponding transfer characteristics measured at $V_{\text{DS}} = 0.1\text{V}$. Although devices with comparable SS generally exhibit comparable $V_{\text{Trans.}}$, distinct recipe-dependent clusters emerge, indicating that fabrication processes simultaneously influence off-state and the transition toward the on-state. This observation suggests that $V_{\text{Trans}}$ provides complementary information beyond conventional transistor metrics and serves as an additional optimization objective for process optimization. Furthermore, the statistical correlation between SS and $V_{\text{Trans}}$ and this recipe-dependent clustering suggests that both metrics are governed by common process-dependent factors affecting gate control efficiency, thus motivating the multi-objective PTCO framework presented in the following sections.

## III. Multi-Objective Ranking of Recipes

## A. Recipes and Process Parameters

Fig. 5 shows the fabrication process flow and the SEM and cross-sectional TEM images of the double-gate 2D monolayer $MoS_2$ FET test vehicle used in this study. Table 1 summarizes the 6 gate stack recipes included in the experimental dataset. The selected process variables represent the primary gate stack parameters intentionally varied across the recipes, while all remaining fabrication conditions were kept constant. This controlled experimental design enables the proposed PTCO framework to systematically evaluate the influence of individual process variations on device switching behavior and identify the most promising fabrication recipes.

## B. Multi-Objective Ranking

Gate-stack optimization requires balancing switching efficiency, drive current and transition behavior. Accordingly, PTCO evaluates each fabrication recipe as a multi-objective optimization problem using statistically extracted device metrics - SS, $I_{ON}$ and $V_{Trans}$ from experimentally measured transfer characteristics as illustrated in Fig. 6(a,b). Fig. 7(a,b) shows the Gaussian-fitted distributions of these three metrics for the six gate-stack recipes. To account for device-to-device variability, the mean and standard deviation of each metric were used in a multi-criteria decision-making framework, with each objective represented as a triangular fuzzy number centered at the recipe-level median and bounded by an uncertainty term scaled by the number of measured devices for that recipe. A Fuzzy TOPSIS framework [4] was then used to quantify each recipe's relative closeness to the ideal solution. SS and $V_{Trans}$ were treated as minimization objectives with weights of 30% each, while $I_{ON}$ was treated as a maximization objective with a weight of 40%, balancing steep subthreshold switching, sharp off-to-on transition and high drive current. The resulting defuzzified scores as shown in Fig. 7(c) provide a variability aware quantitative ranking where larger values indicate more favorable overall trade-offs among the three metrics.

# IV. Machine Learning Prediction and Experimental Validation

## A. Predictive Modeling for Unexplored Recipes

Fig. 8(a) illustrates the PTCO prediction workflow, which links gate stack process parameters to device-level performance metrics across all 1,152 possible recipe combinations. The input feature space consisted of 7 key fabrication parameters while the prediction targets were the statistical descriptors extracted from the device characteristics - SS, $V_{Trans}$ and $I_{ON}$. Due to a limited number of experimentally tested recipes were available, GPR [5] was selected for its strong performance on small datasets, its ability to capture nonlinear relationships, and its inherent prediction uncertainty. For each target metric, GPR models were optimized and benchmarked, after which the best-performing model and its optimal parameters were selected. Graph-kernel GPR was used for SS and $V_{Trans}$, while heteroscedastic ARD-GPR was used for $I_{ON}$. As shown in Fig. 8(b), unexplored candidate recipes were generated by enumerating all combinations of the experimentally explored process parameter levels, expanding the process space beyond the fabricated recipes [6]. PTCO then predicts the expected device performance for each candidate recipe, enabling a data-driven route to evaluate untested recipes and identify promising fabrication conditions with improved multi-objective device performance without exhaustive experimentation.

## B. Ranking and Experimental Validation

The device performance component was evaluated using the Fuzzy TOPSIS score described previously. Because prediction reliability decreases as candidate recipes deviate further from experimentally explored process conditions, an additional confidence score was introduced based on the GPR prediction uncertainty and the distance between each candidate recipe and its nearest experimentally fabricated recipe. As illustrated in Fig. 8(c), the final PTCO ranking combines the predicted device performance score and the confidence score with 50-50% weighting, balancing expected performance against prediction reliability. Experimental validation was performed by fabricating a previously unexplored candidate recipe and comparing its measured device performance and PTCO ranking with the ML prediction. Fig. 9 shows that the experimentally measured ranking (red square) agrees closely with the ML prediction (red star), validating the proposed PTCO framework. Notably, although Recipe 6 employing a Ti TG metal ranked among the lowest performing recipes, PTCO predicted that modifying only the TG oxide deposition temperature would substantially improve its ranking. The experimentally validated Recipe 7 (Table 2) confirms this prediction, demonstrating that interactions among process parameters, rather than any individual parameter alone, govern device performance.

# V. Conclusion

This work demonstrates the first ML-enabled PTCO framework for emerging 2D TMD FETs by integrating automated metric extraction, multi-objective ranking, prediction of unexplored recipes and experimental validation. The introduced transition voltage metric, $V_{Trans}$ complements conventional transistor metrics and enables more comprehensive process performance optimization. Close agreement between the ML predicted and experimentally validated rankings shows that PTCO can identify promising recipes from limited experimental data by capturing coupled process-parameter interactions and enabling recipe-level rather than individual-parameter optimization. To maximize prediction confidence across the full process space, formal design of experiments can be used to identify, fabricate, and measure a broader recipe set for expanded model training. As additional experimental data become available, the PTCO framework can be iteratively retrained to progressively build a self-evolving platform for 2D FET process optimization.

Acknowledgment— This work was funded by a joint MOU between the Indiana Economic Development Corporation, Purdue University and IMEC with in-kind support from the Applied Research Institute and IMEC IIAP Exploratory Logic Program. We thank V. Lunardelli, R. Gafiteanu, and L. Larcher from Applied Materials and acknowledge support from the Ginestra™ Academic Program.

## Variation in 2D Devices and Impact of Device Processing, Material Defects and Trap Charges

Figure 1. (a) Comparison of 2D $MoS_2$ devices with uncontrolled variability with a Si SOI NMOS [1] (b) Schematic of 2D $MoS_2$ FET (c) Ginestra™ simulations showing degraded SS, a broadened off-to-on transition and lower on-current resulting from various defects at various locations.

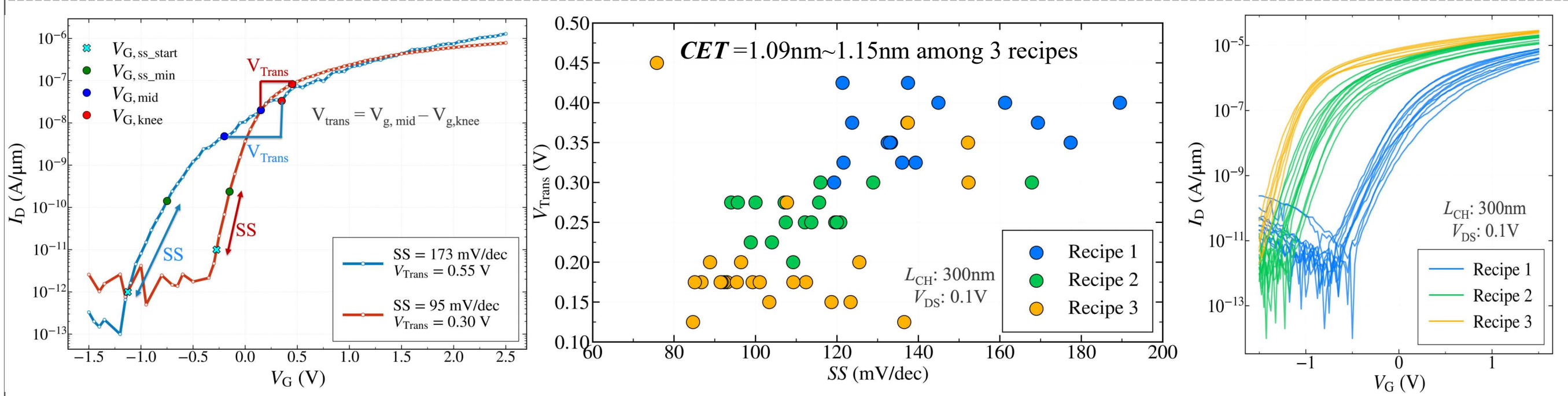

Figure 2. Two representative I-V curves illustrating the identification of key voltage points.

Figure 3. Comparison of extracted transition voltage and SS for three device processing recipes with similar CET.

Figure 4. I-V characteristics of the three recipes used in the $V_{Trans}$-SS extraction.

## Gate Stack Recipes, Fabrication Process Flow, Device Images and Performance Metric Extraction

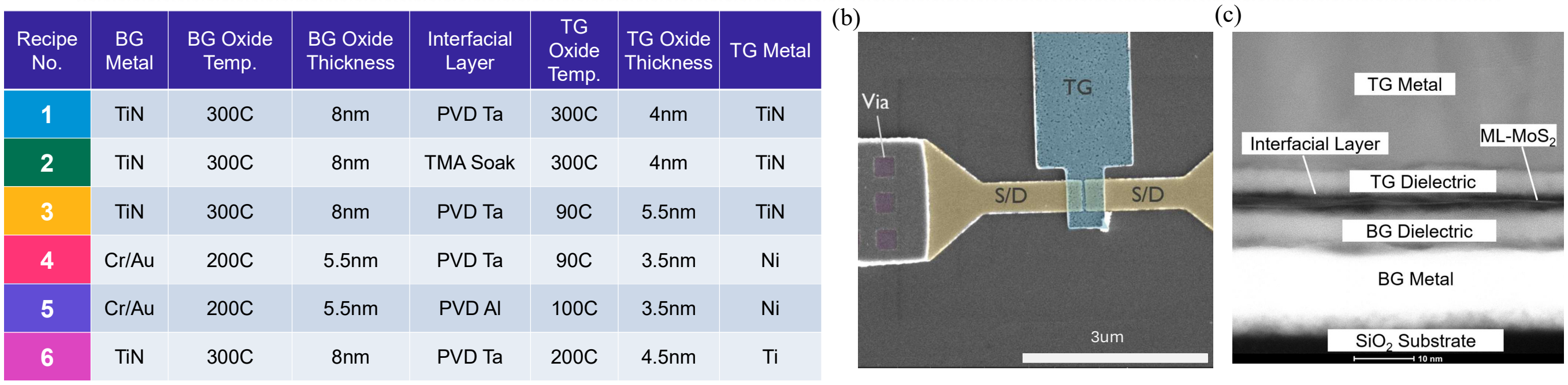

| Recipe No. | BG Metal | BG Oxide Temp. | BG Oxide Thickness | Interfacial Layer | TG Oxide Temp. | TG Oxide Thickness | TG Metal |
|---|---|---|---|---|---|---|---|
| 1 | TiN | 300C | 8nm | PVD Ta | 300C | 4nm | TiN |
| 2 | TiN | 300C | 8nm | TMA Soak | 300C | 4nm | TiN |
| 3 | TiN | 300C | 8nm | PVD Ta | 90C | 5.5nm | TiN |
| 4 | Cr/Au | 200C | 5.5nm | PVD Ta | 90C | 3.5nm | Ni |
| 5 | Cr/Au | 200C | 5.5nm | PVD Al | 100C | 3.5nm | Ni |
| 6 | TiN | 300C | 8nm | PVD Ta | 200C | 4.5nm | Ti |

Table 1. Gate stack recipes defined by 7 varying process parameters for the bottom-gate (BG) and top-gate (TG) stacks

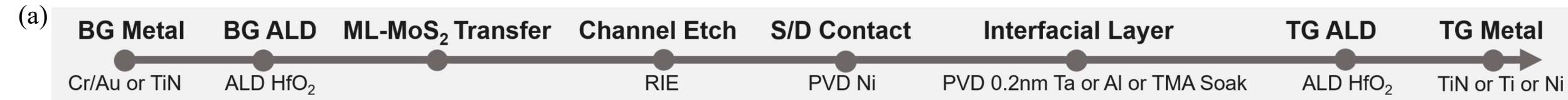

Figure 5. (a) Device fabrication process flow (b) SEM image, and (c) Cross-sectional STEM image of a representative device.

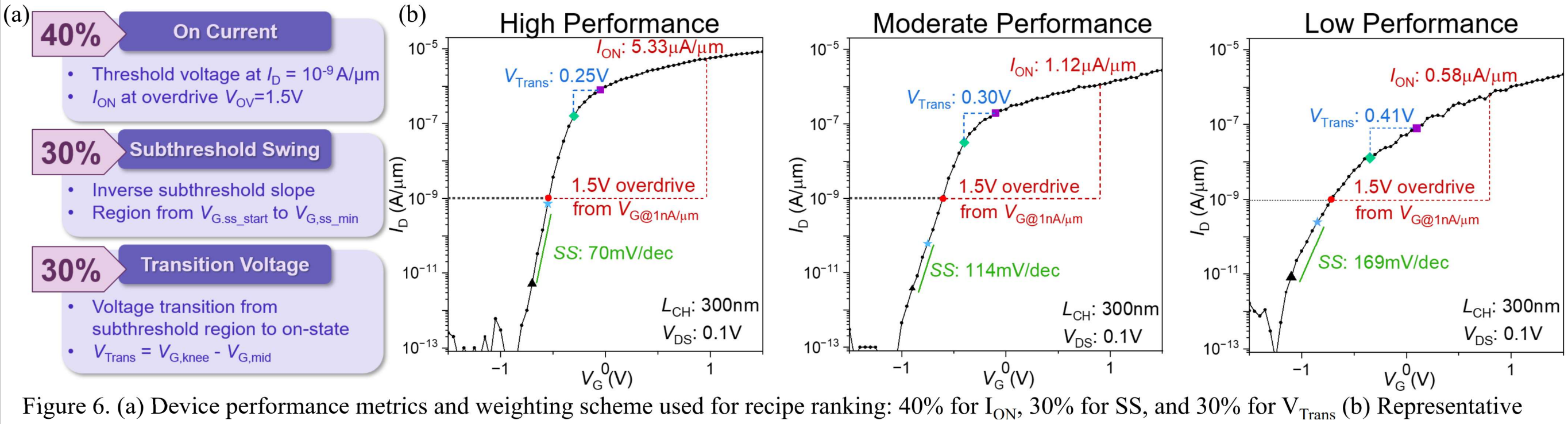

Figure 6. (a) Device performance metrics and weighting scheme used for recipe ranking: 40% for $I_{ON}$, 30% for SS, and 30% for $V_{Trans}$ (b) Representative high-, moderate-, low-performance devices with the extracted performance metrics.

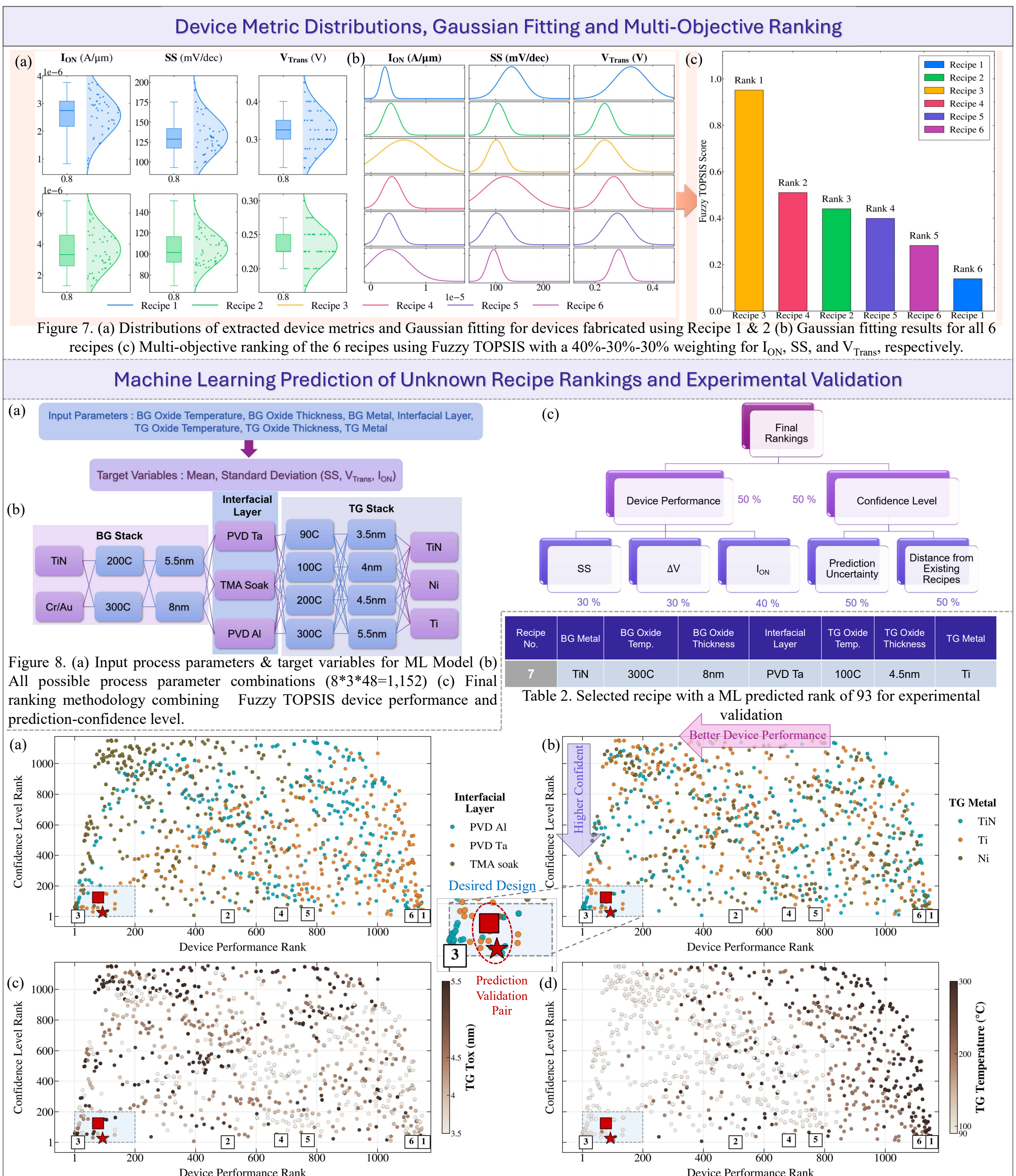

Figure 7. (a) Distributions of extracted device metrics and Gaussian fitting for devices fabricated using Recipe 1 & 2 (b) Gaussian fitting results for all 6 recipes (c) Multi-objective ranking of the 6 recipes using Fuzzy TOPSIS with a 40%-30%-30% weighting for $I_{ON}$, SS, and $V_{Trans}$, respectively.

Figure 8. (a) Input process parameters & target variables for ML Model (b) All possible process parameter combinations (8*3*48=1,152) (c) Final ranking methodology combining Fuzzy TOPSIS device performance and prediction-confidence level.

| Recipe No. | BG Metal | BG Oxide Temp. | BG Oxide Thickness | Interfacial Layer | TG Oxide Temp. | TG Oxide Thickness | TG Metal |
|---|---|---|---|---|---|---|---|
| 7 | TiN | 300C | 8nm | PVD Ta | 100C | 4.5nm | Ti |

Table 2. Selected recipe with a ML predicted rank of 93 for experimental validation

Figure 9. ML predicted conference level vs. device performance ranking for all 1,152 possible process recipes. The 6 experimentally fabricated recipes are highlighted and exhibit high confidence because they are included in the training dataset. Recipe 7's prediction (red star), having a rank of 93 with a confidence level rank of 25, was selected for experimental validation. Its experimentally measured performance is shown as the red square. Colors indicate the effects of different process parameters on the predicted rankings: (a) interfacial layer (b) TG metal (c) TG dielectric thickness (d) TG dielectric deposition temperature. Although individual parameter trends (e.g., lower TG deposition temperature) may appear favorable, device performance is determined by the combined effects of all process parameters. Therefore, optimization should be performed at the recipe level rather than by individual process parameters.